\documentclass[letterpaper, 10 pt, conference]{ieeeconf}  
\usepackage{cite}

\IEEEoverridecommandlockouts                              

\title{\LARGE \bf
Tuning of \lq Constant in gain Lead in phase (CgLp)\rq\, Reset Controller using higher-order sinusoidal input describing function (HOSIDF)
}

\author{Xiaojun Hou$^{1}$, Ali Ahmadi Dastjerdi$^{1}$, Niranjan Saikumar$^{1}$ and S. Hassan Hosseinnia$^{1}$
\thanks{$^{1}$Ali Ahmadi Dastjerdi, Xiaojun Hou, Niranjan Saikumar and S. Hassan Hosseinnia are with  Department of Precision and Microsystems Engineering, Delft University of Technology, 2628CD Delft, The Netherlands
        {\tt\small A.AhmadiDastjerdi@tudelft.nl, X.Hou@student.tudelft.nl, N.Saikumar@tudelft.nl, S.H.HosseinNiaKani@tudelft.nl}}%
}
\usepackage{colortbl}
\usepackage{graphicx}  
\usepackage{subcaption}
\usepackage{amsmath,bm}
\usepackage{mathtools}
\usepackage{amsfonts}
\usepackage {extarrows}
\usepackage{pgfplots}
\usepackage{amssymb}
\usepackage{siunitx}
\usepackage{mathptmx}
\usepackage{anyfontsize}

\usepackage{cancel}
\usepackage{scalerel}
\usepackage{float}
\usepackage{tikz} 
\usepackage{multirow}
\usetikzlibrary{calc,patterns,arrows,shapes.arrows,intersections}
\begin{document}
\maketitle
\thispagestyle{empty}
\pagestyle{empty}

\begin{abstract}
Due to development of technology, linear controllers cannot satisfy requirements of high-tech industry. One solution is using nonlinear controllers such as reset elements to overcome this big barrier. In literature, the Constant in gain Lead in phase (CgLp) compensator is a novel reset element developed to overcome the inherent linear controller limitations. However, a tuning guideline for these controllers has not been proposed so far. In this paper, a recently developed method named higher-order sinusoidal input describing function (HOSIDF), which gives deeper insight into the frequency behaviour of non-linear controllers compared to sinusoidal input describing function (DF), is used to obtain a straight-forward tuning method for CgLp compensators. In this respect, comparative analyses on tracking performance of these compensators are carried out. Based on these analyses, tuning guidelines for CgLp compensators are developed and validated on a high-tech precision positioning stage. The results show the effectiveness of the developed tuning method.
\end{abstract}
\section{INTRODUCTION}

Development of the high-tech industry has pushed the requirements of motion control applications to extremes regarding precision, stability, and speed. Thus, Proportional Integral Derivative (PID), which has been widely used in industry for its ease of implementation and simple structure, cannot satisfy these requirements due to fundamental limitations - waterbed effect \cite{1,schmidt2014design}. To overcome these limitations, researchers have turned to nonlinear controllers such as reset controllers \cite{9,10,hunnekens2014synthesis,van2018hybrid,11,13}.

In 1958, Clegg \cite{2}  proposed the first reset element which resets the state of an integrator to zero when its input crosses the zero point. Besides Clegg Integrator (CI), other reset configurations have been developed to provide more design freedom and applicability: Generalized First Order Reset Element (GFORE) \cite{4} and Generalized Second Order Reset Element (GSORE) \cite{5}. Apart from zero error crossing condition, other conditions like reset band \cite{barreiro2014reset,banos2014tuning} and fixed reset instants \cite{8.5} have also been studied. Moreover, there are several techniques which are proposed to soften nonlinearities of reset controllers such as Partial Reset and PI+CI approaches \cite{banos2011reset}. 

Describing Function (DF) tool, which considers only the first harmonic of the output of the controller for a sinusoidal input shows that the gain slope of CI is the same as the linear integrator while it produces $52^\circ$ less phase lag than the latter. This phase lead advantage of reset controllers has been used to introduce new phase compensators \cite{valerio2019reset,saikumar2019constant,van2018hybrid}. N. Saikumar et. al. \cite{saikumar2019constant} proposed a novel reset element termed 'Constant in gain Lead in phase (CgLp)' which produces broadband  phase lead while maintaining constant gain. This compensator is made by combining the GFORE/GSORE with the first/second order linear lead filter. 

As a result of the design flexibility of CgLp compensators, various combinations of tuning parameters could be used to provide the same open-loop gain behaviour and  phase compensation at the crossover frequency based on the DF analysis for these compensators. However, it was seen that the improvement expected through describing function analyses was not achieved in some cases \cite{17}. Hence, DF analysis is insufficient to perform frequency analyses for reset elements especially when precision applications are considered. Recently, Nuij \cite{18} has extended the DF method to higher order sinusoidal input describing functions (HOSIDF) for the frequency analyses of non-linearities such as backlash, friction, etc. With this tool, Heinen \cite{19} extended HOSIDF for reset controllers and opened the possibility of more accurate analyses of reset controllers. The objective of this paper is to use HOSIDF to develop a tuning guideline for CgLp compensators.

The structure of the paper is as follows. Section \ref{sec:2} gives preliminaries on reset controllers. In Section \ref{sec:3}, the tuning method is derived based on the analyses of DF and HOSIDF using the simulation results as basis for time domain performance. Then, Section \ref{sec:4} presents the experiment verification and, conclusions and remarks for further study are provided in Section \ref{sec:5}.
\section{PRELIMINARIES}\label{sec:2}
\subsection{Describing function (DF) and higher order sinusoidal input describing function (HOSIDF)} 
The state-space representation of reset controllers is:
\begin{equation}
\left \{
\begin{array} 
{ l l } { \dot { x }  ( t ) = A_rx ( t ) + B_re ( t ) } & { \text { if } e ( t ) \neq 0 } \\ { x\left( t ^ { + } \right) = A _\rho x ( t ) } & { \text { if } e ( t ) = 0 } \\ { u ( t ) = C_rx ( t ) + D_r e ( t ) } 
\end{array}
\right.
 \label{reset}
\end{equation}
where $A_r$, $B_r$, $C_r$ and $D_r$ are the state matrices of the base linear system, $e(t)$ and $u(t)$ are the error input and control input, respectively. In addition, the resetting matrix $A_\rho$ determines states' value after reset action by which the non-linearity of reset systems can be tuned. The sinusoidal input DF of reset systems (\ref{reset}) is given in \cite{6} as:
 \begin{equation}
 {G_{DF} ( j \omega ) = C_r  \left( j \omega I - A_r \right) ^ { - 1 } B_r  \left( I + j \Theta _ { D } ( \omega ) \right) + D_r }
\end{equation}
 where $\Theta _ D $ is:
\begin{equation}\label{E-02-03-01}
\resizebox{\hsize}{!}{$
\Theta_D(\omega)=\dfrac{-2\omega^2}{\pi}(I+e^{\frac{\pi A_r}{\omega}})\big((I+A_\rho e^{\frac{\pi A_r}{\omega}})^{-1}A_\rho(I+e^{\frac{\pi A_r}{\omega}})-I\big)(\omega^2I+A_r^2)^{-1}
$}
\end{equation}
To include higher order harmonics and obtain a more reliable frequency description of reset systems, HOSIDF is obtained in \cite{19} as:
\begin{equation}
\resizebox{\hsize}{!}{$
      G(j\omega , n) = \left\{ \begin{array} { c c } { C_r  ( j \omega n I - A_r ) ^ { - 1 } j \Theta _ { D } ( \omega ) B_r } & { \text { for odd } n \geq 2 } \\ { 0 } & { \text { for even } n \geq 2 } \end{array} \right.
      \label{HOSIDF}
$}
\end{equation}
where $n$ is the order of harmonics. 

\begin{figure}
    \centering
    \begin{tikzpicture}
    \node[anchor=south west,inner sep=0] at (0,0) {\includegraphics[width=0.8\columnwidth]{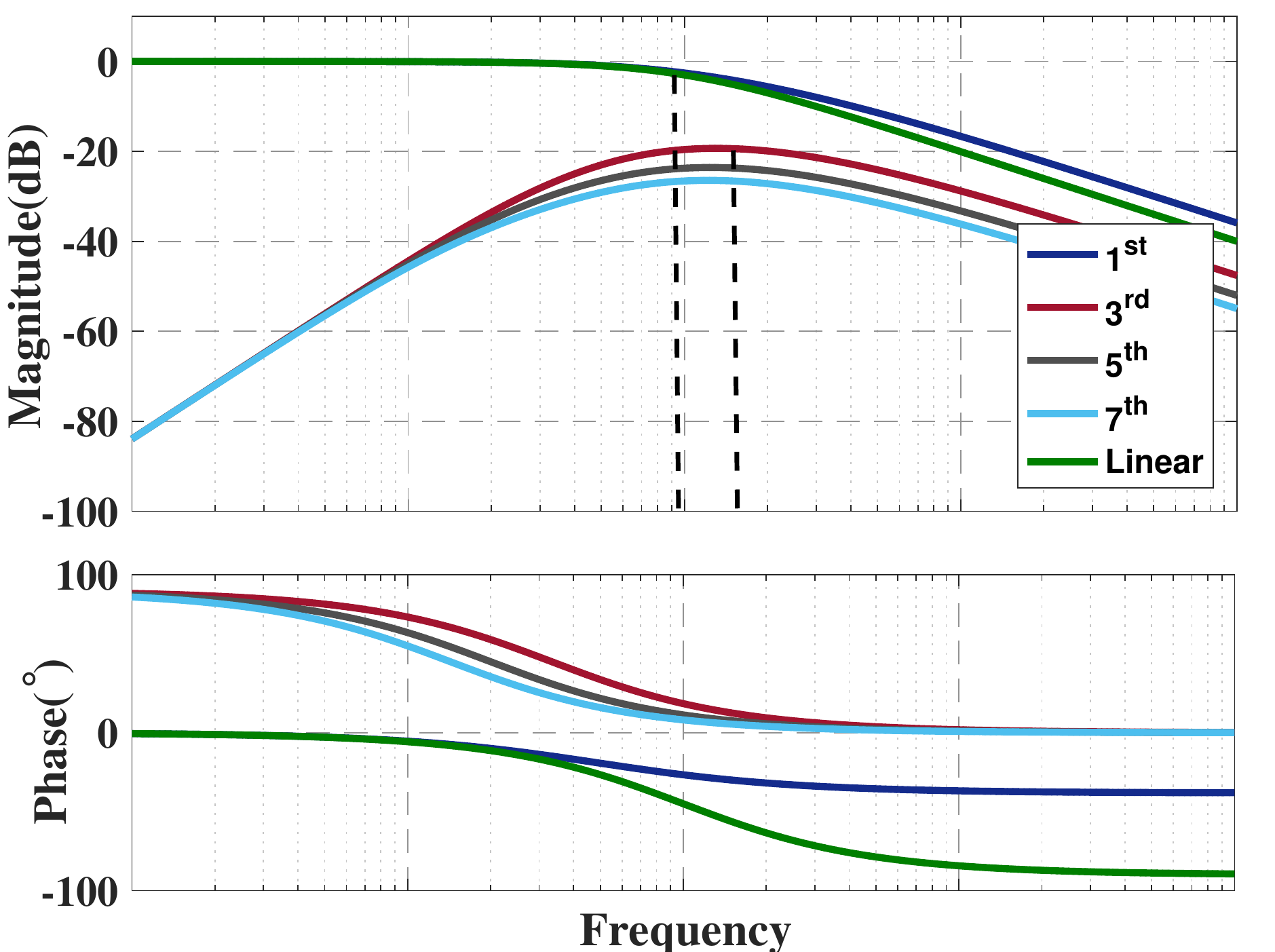}};
\draw (3.65,2.25) node [scale=0.75, rotate =0]  {$\omega_{r\alpha}$};
\draw (4.1,2.25) node [scale=0.75]  {$\omega_p$};
\end{tikzpicture}
\caption{Frequency behaviour of FORE including its linear base, DF, and higher order harmonics}
 \label{2}
\end{figure}

Now consider a GFORE which is represented as
\begin{equation}\label{E-01}
\text{GFORE} =\dfrac { 1 } {\cancelto{A_\rho}   {\dfrac{s }{\omega _ { r \alpha}}   + 1}} 
\end{equation}
where $\omega_{r\alpha}$ is the corner frequency of GFORE and $\gamma$ determines reset values after reset action $A_\rho=\gamma I_{n_r\times n_r}$.
Based on the above relations, harmonics of the GFORE (with $\gamma=0$) along with the frequency response of its base linear system (first order filter) are shown in Fig. \ref{2}. Based on the first harmonic, it can be seen that the reduction of phase lag is obtained without significant change of gain magnitude. Based on (\ref{HOSIDF}), since the magnitude of higher order harmonics decreases with increasing order $(n)$, we only use the third harmonics to analyze the effect of the higher order harmonics on closed-loop performances. Denote $\omega_p$ as frequency at which the magnitude of higher order harmonics reach the peak and $M_p$ as the peak magnitude of $3^{rd}$ harmonic.
\subsection{Pseudo-sensitivity for reset systems}\label{meitric}
In linear systems, the sensitivity function from reference signal $r(t)$ to error $e(t)$ can be calculated by 
\begin{equation}\label{linearsnesitivity}
S(s)=\frac{e}{r}= \frac { 1 } { 1 + G(s) C(s)} 
\end{equation} 
where $G(s)$ and $C(s)$ are the transfer function of the plant and controller, respectively. This transfer function indicates the ability of the system to precisely track the reference signal. 

For non-linear controllers, $C(s)$ can be substituted with DF of the controller to analyze tracking performance. However, it is not accurate enough to predict the precision of tracking performance since higher order harmonics are neglected. To obtain a better indicator for reset systems, a pseudo-sensitivity function ($S_{\infty}$) for a sinusoidal reference $r=r_0\sin(\omega t)$ is defined as:
 \begin{equation}\label{PS}
\forall \omega :\ |S_{\infty}(\omega)|=\frac{\max \limits_{ t \geq t _ { s s } } ( | e ( t ) | )}{r_0 }  = \frac{\max \limits_ { t \geq t _ { s s } }( | r ( t ) - y ( t ) | )}{r_0}
\end{equation}
where $y(t)$ is the system output, and $t_{ss}$ indicates the time after which the system reaches its steady-state output.
\subsection{Constant in gain Lead in phase (CgLp) compensators}
\begin{figure*}
  \centering
  \begin{subfigure}{\columnwidth}
    \centering
   \begin{tikzpicture}
    \node[anchor=south west,inner sep=0] at (0,0) {\includegraphics[width=0.7\linewidth]{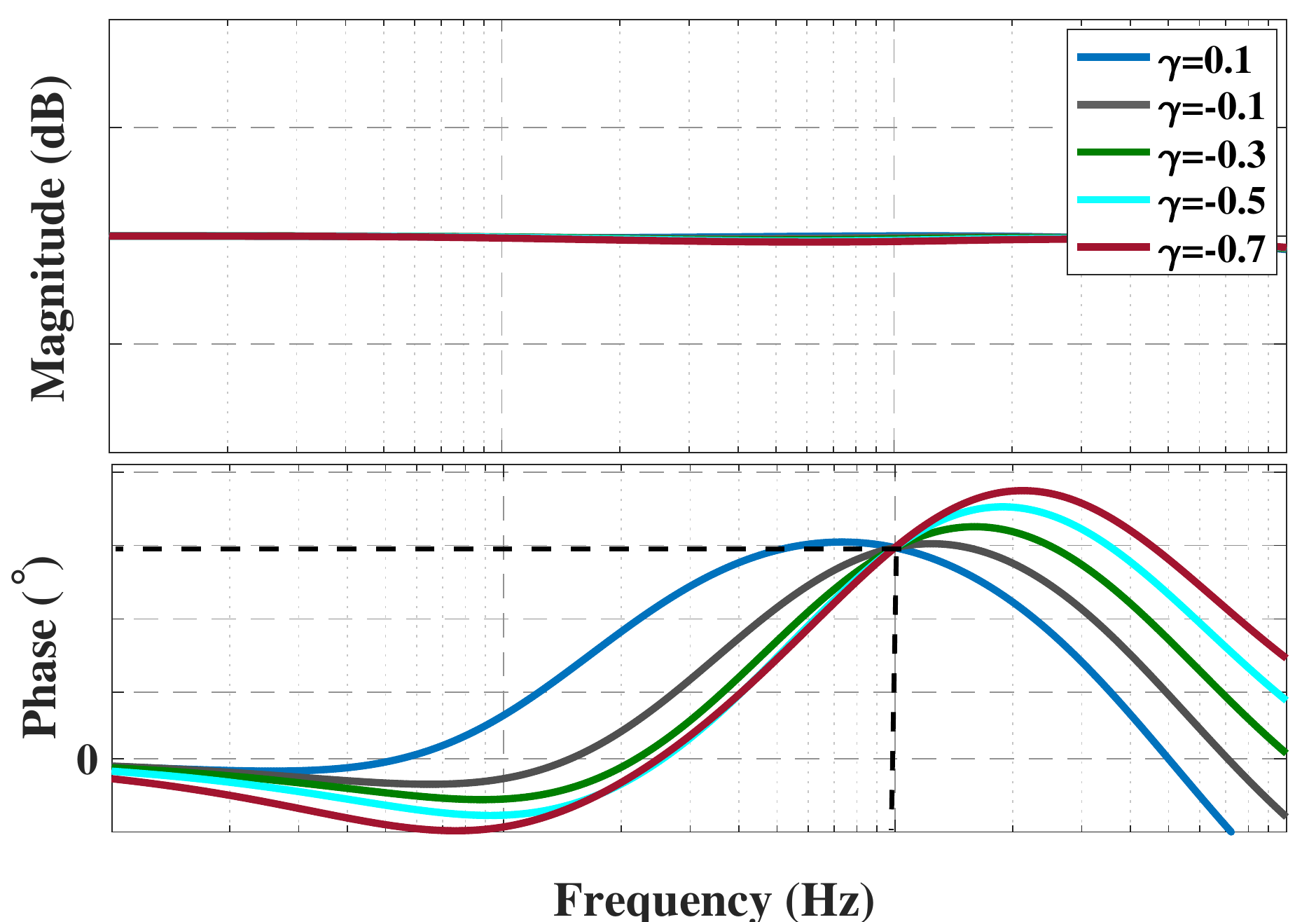}};
\draw (4.15,0.3) node [scale=0.8]  {$\omega_c$};
\draw (0.44,1.7) node [scale=0.8]  {$\theta$};
\end{tikzpicture}
\caption{DF of CgLp}
 \label{DFomegar}
     \end{subfigure}
\hfill
\begin{subfigure}{\columnwidth}
    \centering
     \begin{tikzpicture}
    \node[anchor=south west,inner sep=0] at (0,0) {\includegraphics[width=0.7\columnwidth]{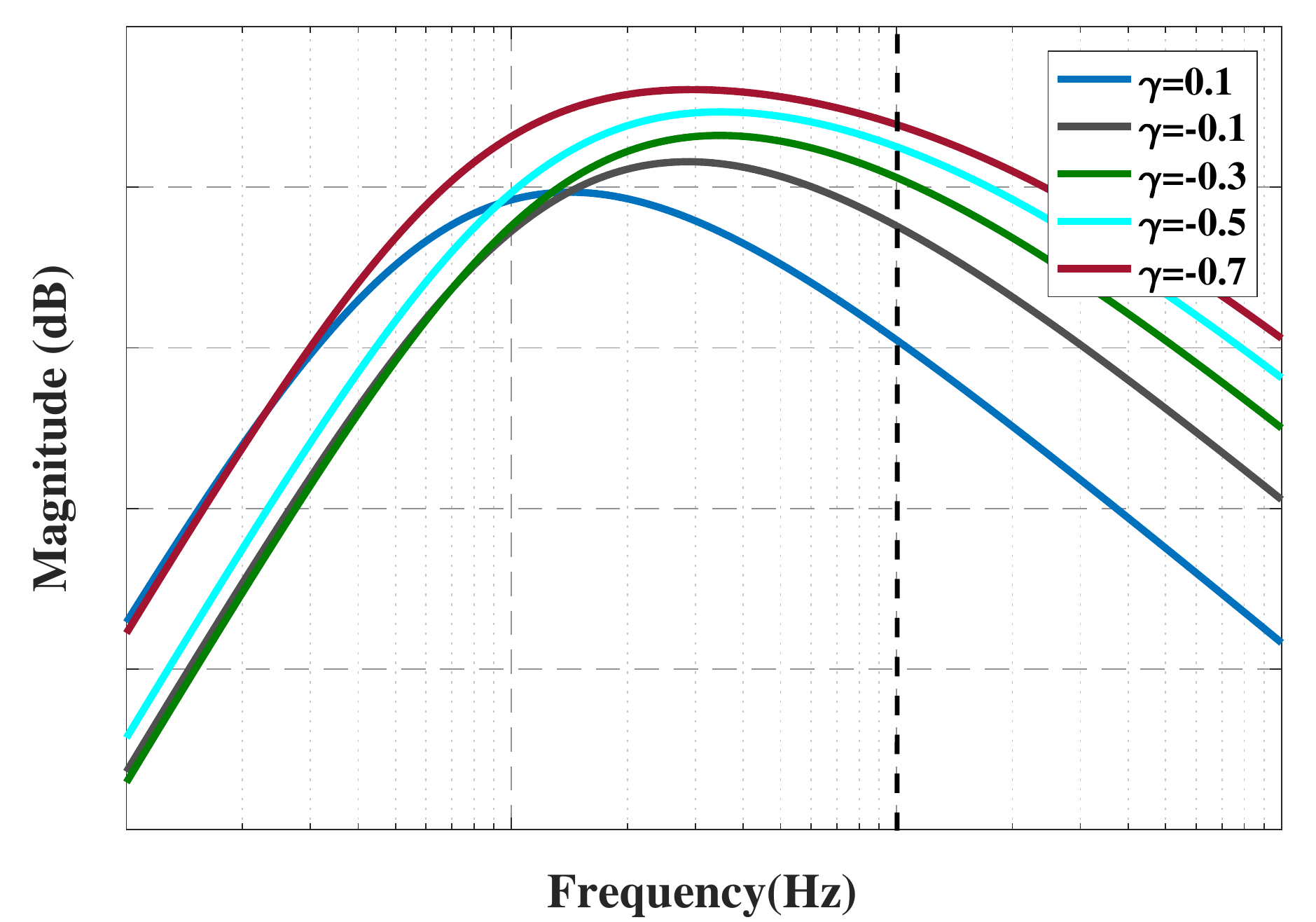}};
\draw (4.15,0.3) node [scale=0.8]  {$\omega_c$};
\end{tikzpicture}
    \caption{$3^{rd}$ harmonic of the reset element of CgLp}
    \label{mp}
  \end{subfigure}  
  \caption{The Frequency behaviour of CgLp element including DF and $3^{rd}$ order harmonic of reset element}  
  \label{CG}
\end{figure*}
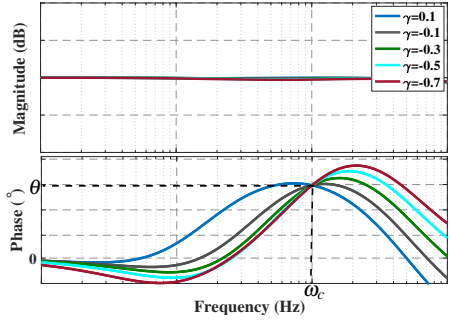
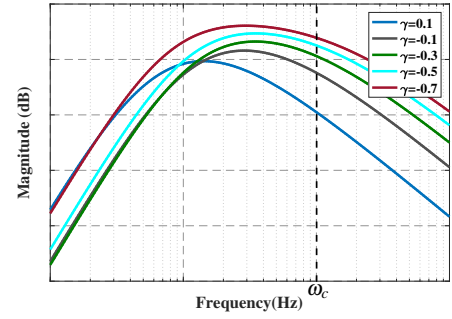

In \cite{saikumar2019constant}, CgLp is introduced as a phase compensator by combining GFORE or GSORE with a corresponding order of lead filter. It was proven that CgLp can be used advantageously to overcome the fundamental limitations of linear controllers. The first order CgLp is defined as follows:
\begin{equation}\label{E-01}
  C_{CgLp1}(s) =\dfrac { 1 } {\cancelto{A_\rho}   {\dfrac{s }{\omega _ { r \alpha}}   + 1}} \quad  \left(\dfrac { \dfrac{s}{ \omega _ { r } }  + 1} { \dfrac{s}{\omega_t}+1 }\right) 
\end{equation}
in which $\omega_r$ and $\omega_t$ are the  starting and taming frequencies of linear lead filter. The corner frequency of the GFORE is set $\omega_{r\alpha}= \frac{\omega_{r}}{\alpha}$ where $\alpha$ is the correction factor accounting for the shift of the corner frequency of GFORE to ensure constant gain of CgLp elements. The values of $\alpha$ with respect to $\gamma$ are listed in \cite{saikumar2019constant}. Similarly, the second order CgLp can be defined as:
\begin{equation}\label{E-02}
\resizebox{\hsize}{!}{$
   C_{CgLp2}(s)=  \dfrac {1 } {\cancelto{A_\rho}{( \dfrac{s}{\omega_{r \alpha}} ) ^ { 2 } +  2 s \dfrac{\beta_{r\alpha} }{ \omega _ {  r\alpha}}  + 1}}\quad \left(\dfrac {  \left( \dfrac{s}{ \omega_{r }} \right) ^ { 2 } + 2 s\dfrac{ \beta_{r}} { \omega _ {r} } + 1} {\left( \dfrac{s}{\omega_t}+1 \right) ^ { 2 } }\right) 
$}
\end{equation}
where $\omega_{r\alpha}=\frac{\omega_{r}}{\alpha_1} $ and    $\beta_{r\alpha}= \frac{\beta_{r}}{\alpha_2}$ are the corner frequency and damping ratio of the reset element, respectively. $\alpha_1$ and $\alpha_2$ (Table \ref{2ndCgLp}) are correction factors considering the shift of corner frequency and adjustment of damping ratio to guarantee constant gain of second order CgLp elements ($\beta_{r \alpha}$ is considered as $1$ in this paper).
\begin{table*}[t!]
\centering
\caption{Correction factors $\alpha_1$ and $\alpha_2$ of second order CgLp}
\resizebox{\textwidth}{!}{
\begin{tabular}{|c|c|c|c|c|c|c|c|c|c|c|c|c|c|c|c|c|c|c|c|}
\hline
$\bm{\gamma}$                           & -0.9                      & -0.8                      & -0.7                      & -0.6                      & -0.5                      & -0.4                      & -0.3                      & -0.2                      & -0.1                      & 0                         & 0.1                       & 0.2                       & 0.3                       & 0.4                       & 0.5                       & 0.6                       & 0.7                       & 0.8                       & 0.9                       \\ \hline
$\bm{\alpha_1}$                       & 30.09                     & 14.11                     & 8.66                      & 5.89                      & 4.23                      & 3.11                      & 2.43                      & 1.92                      & 1.52                      & 1.23                      & 1.03                      & 0.93                      & 0.89                      & 0.90                      & 0.92                      & 0.94                      & 0.96                      & 0.98                      & 0.99                      \\ \hline
\multicolumn{1}{|l|}{$\bm{\alpha_2}$} & \multicolumn{1}{l|}{3.28} & \multicolumn{1}{l|}{3.20} & \multicolumn{1}{l|}{3.01} & \multicolumn{1}{l|}{2.76} & \multicolumn{1}{l|}{2.49} & \multicolumn{1}{l|}{2.21} & \multicolumn{1}{l|}{2.10} & \multicolumn{1}{l|}{1.91} & \multicolumn{1}{l|}{1.63} & \multicolumn{1}{l|}{1.36} & \multicolumn{1}{l|}{1.14} & \multicolumn{1}{l|}{1.02} & \multicolumn{1}{l|}{1.00} & \multicolumn{1}{l|}{1.03} & \multicolumn{1}{l|}{1.06} & \multicolumn{1}{l|}{1.07} & \multicolumn{1}{l|}{1.07} & \multicolumn{1}{l|}{1.05} & \multicolumn{1}{l|}{1.03} \\ \hline
\end{tabular}}
\label{2ndCgLp}
\end{table*}
The phase compensation provided by CgLp elements at a pre-determined cross-over frequency $\omega_c$ of a system can be defined as $\theta(\omega_r, \gamma)$. The mapping of $\theta$ to $\omega_r$ and $\gamma$ is one to many; hence, no unique solution exists for $\gamma$ and $\omega_r$. Figure \ref{CG} shows DF and HOSIDF behaviour of CgLp configurations which provide the same amount of phase compensation at $\omega_c$. In this paper, we will analyze the $3^{rd}$ harmonic of the constituent reset element (GFORE or GSORE) instead of the entire CgLp in order to find a relation between performance and higher order harmonics. As shown in Fig. \ref{mp}, it is seen that the $3^{rd}$ harmonic of the different configurations are quite different even though their first harmonics are very similar. It is observed that the $M_p$ increases with decreasing $\gamma$ ($M_p$ is only depended on $\gamma$). 

To summarize, it is desired to tune CgLp parameters $\gamma$ and $\omega_r$ such that the required phase $(\theta)$ is achieved at the cross-over frequency $\omega_c$ and the negative effects of higher order harmonics are minimized.
\section{Tuning Guidelines}\label{sec:3}
This section presents the process of developing CgLp tuning guidelines in simulation. Once several CgLp elements are designed to provide the same phase compensation $\theta$, it is assumed that the case with optimal tracking performance is affected the least negatively by higher order harmonics. The tracking is evaluated based on the pseudo-sensitivity ($S_{\infty}$). Then, the relation between the open-loop $3^{rd}$ harmonic behaviour and closed-loop tracking performance is postulated, which leads to the tuning guidelines.
\subsection{Design of controllers}
Due to the design flexibility of CgLp configurations, several groups of the first and second order CgLp elements are designed to produce $\theta$ ($20^\circ,\,30^\circ,\,40^\circ,50^\circ$) phase compensation at the cross-over frequency $\omega_c$ by varying $\gamma$ and $\omega_r$. The value of  $\gamma$ is chosen from -0.9 to 0.9 (to guarantee the open-loop convergence\cite{6}) with an increment of 0.1, and $\omega_r=\frac{\omega_c}{b}$ where $b$ is used to obtain the corner frequency of CgLp elements and is determined by $\gamma$ and $\theta$. Furthermore, $\omega_t=5\omega_c$ to reduce lag effect on the phase margin of the controller while attenuating noise of the system.
\begin{table*}[t!]
\centering
 \caption{Tuning parameters for the first (\ref{E-01}) and second (\ref{E-02}) order CgLp configurations}
{
\setlength\arrayrulewidth{1.05pt}
\resizebox{\textwidth}{!}{
\begin{tabular}{|c|c|c|c|c|c|c|c|c|c|c|c|c|c|c|c|c|c|}
\hline
\multicolumn{2}{|c|}{\multirow{2}{*}{\begin{tabular}[c]{@{}c@{}}CgLp\\ Type\end{tabular}}}                                        & \multicolumn{4}{c|}{$20^\circ$}                            & \multicolumn{4}{c|}{$30^\circ$}                                                                                                               & \multicolumn{4}{c|}{$40^\circ$}                                                                                                               & \multicolumn{4}{c|}{$50^\circ$}                                                                                                               \\ \cline{3-18} 
\multicolumn{2}{|c|}{}                                                             & $\gamma$ & $b$  & $\dfrac{\omega_c}{\omega_p}$ & $M_p(dB)$ & \multicolumn{1}{l|}{$\gamma$} & \multicolumn{1}{l|}{$b$} & \multicolumn{1}{l|}{$\dfrac{\omega_c}{\omega_p}$} & \multicolumn{1}{l|}{$M_p(dB)$} & \multicolumn{1}{l|}{$\gamma$} & \multicolumn{1}{l|}{$b$} & \multicolumn{1}{l|}{$\dfrac{\omega_c}{\omega_p}$} & \multicolumn{1}{l|}{$M_p(dB)$} & \multicolumn{1}{l|}{$\gamma$} & \multicolumn{1}{l|}{$b$} & \multicolumn{1}{l|}{$\dfrac{\omega_c}{\omega_p}$} & \multicolumn{1}{l|}{$M_p(dB)$} \\ \hline
\multirow{2}{*}{\begin{tabular}[c]{@{}c@{}}First\\ Order\end{tabular}}  & Tracking & -0.3     & 1.34 & 1.94                         & -16.77    & -0.4                          & 1.71                     & 2.81                                              & -16.02                         & -0.5                          & 2.23                     & 4.25                                              & -15.3                          & -0.6                          & 3.07                     & 7.03                                              & -14.62                         \\ \cline{2-18} 
                                                                        & Noise    & 0.3      & 6.41 & 6.11                         & -22.63    & 0.1                           & 6.8                      & 7.05                                              & -20.31                         & 0.0                           & 25.82                    & 28.49                                             & -19.32                         & -0.2                          & 23.78                    & 31.05                                             & -17.56                         \\ \hline
\multirow{2}{*}{\begin{tabular}[c]{@{}c@{}}Second\\ Order\end{tabular}} & Tracking & 0.2      & 0.75 & 1.06                         & -18.59    & 0.2                           & 1.0                      & 1.43                                              & -18.59                         & 0.2                           & 1.35                     & 1.91                                              & -18.59                         & -0.1                          & 1.26                     & 1.48                                              & -15.73                         \\ \cline{2-18} 
                                                                        & Noise    & 0.5      & 1.44 & 2.07                         & -22.74    & 0.4                           & 1.61                     & 2.25                                              & -21.14                         & 0.3                           & 1.39                     & 2.34                                              & -19.77                         & 0.3                           & 3.53                     & 4.63                                              & -19.77                         \\ \hline
\end{tabular}
}}
\label{foretuning}
\end{table*}
\subsection{Closed-loop precision performance}
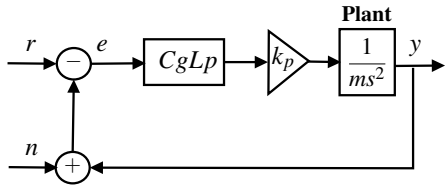
\begin{figure}
  \centering
\resizebox{0.7\columnwidth}{!}{
\tikzset{every picture/.style={line width=0.75pt}} 
\begin{tikzpicture}[x=0.75pt,y=0.75pt,yscale=-1,xscale=1]
\draw  [line width=1.5]  (48.63,58.45) .. controls (48.63,50.84) and (55.3,44.68) .. (63.53,44.68) .. controls (71.76,44.68) and (78.43,50.84) .. (78.43,58.45) .. controls (78.43,66.06) and (71.76,72.22) .. (63.53,72.22) .. controls (55.3,72.22) and (48.63,66.06) .. (48.63,58.45) -- cycle ;
\draw [line width=1.5]    (360.5,62) -- (360.5,150) -- (81.43,149.46) ;
\draw [shift={(77.43,149.45)}, rotate = 360.11] [fill={rgb, 255:red, 0; green, 0; blue, 0 }  ][line width=0.08]  [draw opacity=0] (11.61,-5.58) -- (0,0) -- (11.61,5.58) -- cycle    ;
\draw [line width=1.5]    (5.5,58) -- (44.63,58.41) ;
\draw [shift={(48.63,58.45)}, rotate = 180.6] [fill={rgb, 255:red, 0; green, 0; blue, 0 }  ][line width=0.08]  [draw opacity=0] (11.61,-5.58) -- (0,0) -- (11.61,5.58) -- cycle    ;
\draw [line width=1.5]    (344.77,60.37) -- (384.5,60.03) ;
\draw [shift={(388.5,60)}, rotate = 539.52] [fill={rgb, 255:red, 0; green, 0; blue, 0 }  ][line width=0.08]  [draw opacity=0] (11.61,-5.58) -- (0,0) -- (11.61,5.58) -- cycle    ;
\draw [line width=1.5]    (268.49,56.5) -- (293.5,56.93) ;
\draw [shift={(297.5,57)}, rotate = 180.99] [fill={rgb, 255:red, 0; green, 0; blue, 0 }  ][line width=0.08]  [draw opacity=0] (11.61,-5.58) -- (0,0) -- (11.61,5.58) -- cycle    ;
\draw  [line width=1.5]  (296.17,28.09) -- (344.5,28.09) -- (344.5,86) -- (296.17,86) -- cycle ;
\draw [line width=1.5]    (194.5,57) -- (229.5,57) ;
\draw [shift={(233.5,57)}, rotate = 180] [fill={rgb, 255:red, 0; green, 0; blue, 0 }  ][line width=0.08]  [draw opacity=0] (11.61,-5.58) -- (0,0) -- (11.61,5.58) -- cycle    ;
\draw  [line width=1.5]  (47.63,149.45) .. controls (47.63,141.84) and (54.3,135.68) .. (62.53,135.68) .. controls (70.76,135.68) and (77.43,141.84) .. (77.43,149.45) .. controls (77.43,157.06) and (70.76,163.22) .. (62.53,163.22) .. controls (54.3,163.22) and (47.63,157.06) .. (47.63,149.45) -- cycle ;
\draw [line width=1.5]    (62.53,135.68) -- (63.47,76.22) ;
\draw [shift={(63.53,72.22)}, rotate = 450.9] [fill={rgb, 255:red, 0; green, 0; blue, 0 }  ][line width=0.08]  [draw opacity=0] (11.61,-5.58) -- (0,0) -- (11.61,5.58) -- cycle    ;
\draw [line width=1.5]    (6.5,149) -- (45.63,149.41) ;
\draw [shift={(49.63,149.45)}, rotate = 180.6] [fill={rgb, 255:red, 0; green, 0; blue, 0 }  ][line width=0.08]  [draw opacity=0] (11.61,-5.58) -- (0,0) -- (11.61,5.58) -- cycle    ;
\draw  [line width=1.5]  (269.18,56.01) -- (233.88,84.78) -- (234.22,29.88) -- cycle ;
\draw  [line width=1.5]  (125,37) -- (195,37) -- (195,77) -- (125,77) -- cycle ;
\draw [line width=1.5]    (78.43,58.45) -- (122.5,58.04) ;
\draw [shift={(126.5,58)}, rotate = 539.46] [fill={rgb, 255:red, 0; green, 0; blue, 0 }  ][line width=0.08]  [draw opacity=0] (11.61,-5.58) -- (0,0) -- (11.61,5.58) -- cycle    ;

\draw (63.53,56.61) node  [scale=1.2,font=\large]  {$-$};
\draw (25.84,42.76) node  [scale=1.4,font=\large]  {$r$};
\draw (362.62,44.76) node  [scale=1.4,font=\large]  {$y$};
\draw (88.35,42.76) node  [scale=1.4,font=\large]  {$e$};
\draw (320.34,57.04) node  [scale=1.3,font=\large]  {$\dfrac{1}{ms^{2}}$};
\draw (62.53,147.61) node  [scale=1.2,font=\large]  {$+$};
\draw (164,57) node  [scale=1.4,font=\large]  {$CgLp$};
\draw (26.84,133.76) node  [scale=1.4,font=\large]  {$n$};
\draw (320,15) node   [scale=1.2,align=left] {\textbf{{\large {\fontfamily{ptm}\selectfont Plant}}}};
\draw (247,54) node  [scale=1.4,font=\large]  {$k_{p}$};
\end{tikzpicture}
}
    \caption{Block diagram of the closed-loop system including a mass plant controlled by CgLp compensators}
    \label{c1}
  \end{figure}

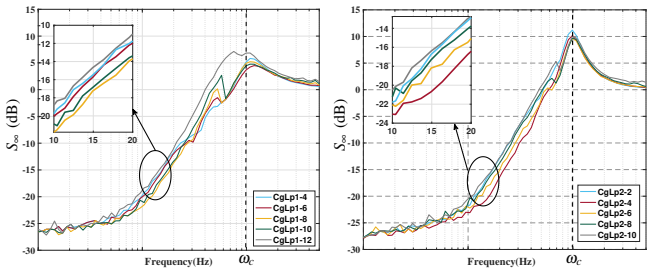
\begin{figure}
  \centering
  \begin{subfigure}{0.49\columnwidth}
    \centering
    \begin{tikzpicture}
    \node[anchor=south west,inner sep=0] at (0,0) {\includegraphics[width=\linewidth]{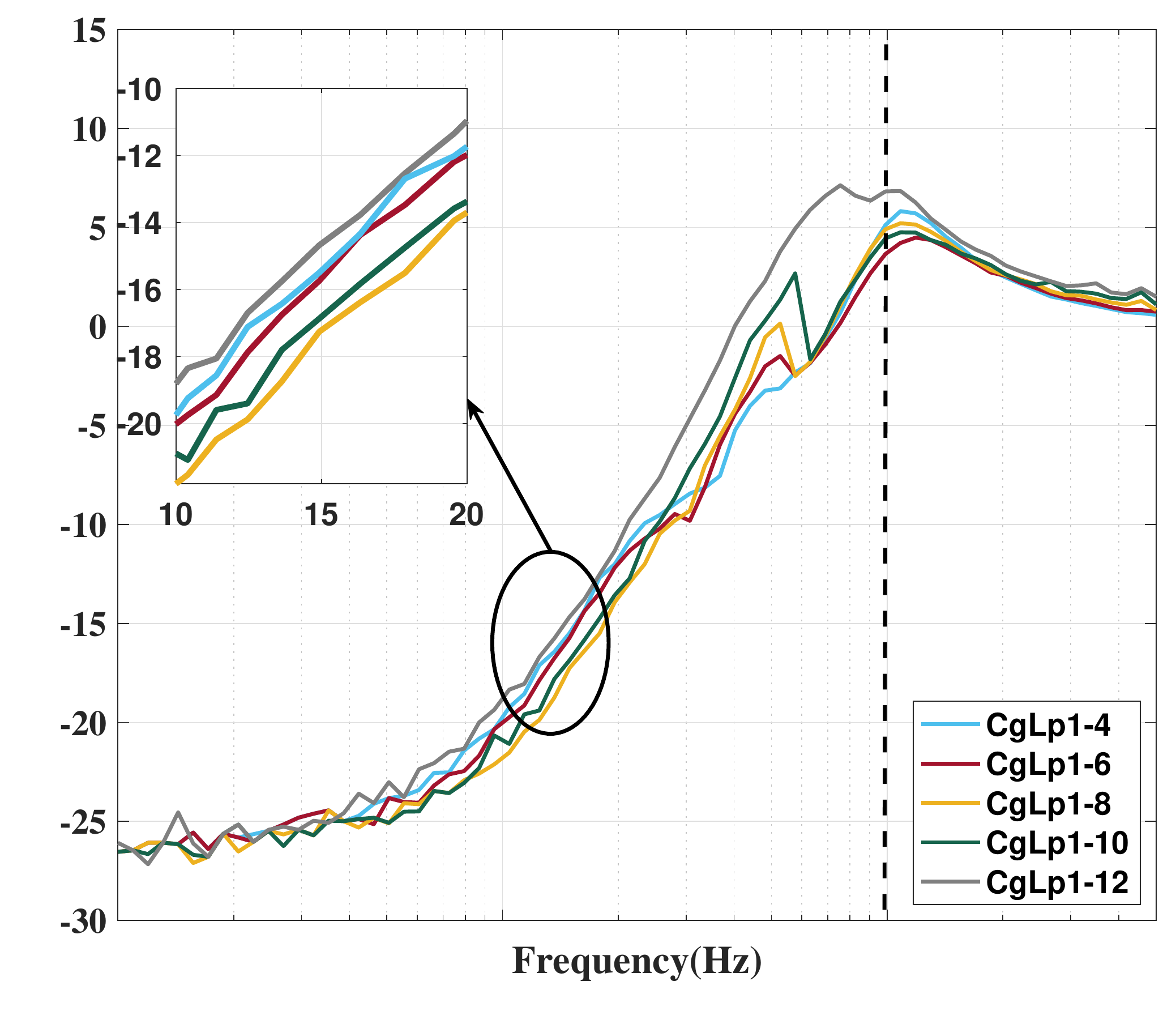}};
\draw (3.2,0.2) node [scale=0.6]  {$\omega_c$};
\draw (0.1,2) node [scale=0.6, rotate=90]  {$S_\infty$ (dB)};
\end{tikzpicture}
    \caption{The first order CgLp}
    \label{subfiga}
  \end{subfigure}
\begin{subfigure}{0.49\columnwidth}
    \centering
    \begin{tikzpicture}
    \node[anchor=south west,inner sep=0] at (0,0) {\includegraphics[width=\linewidth]{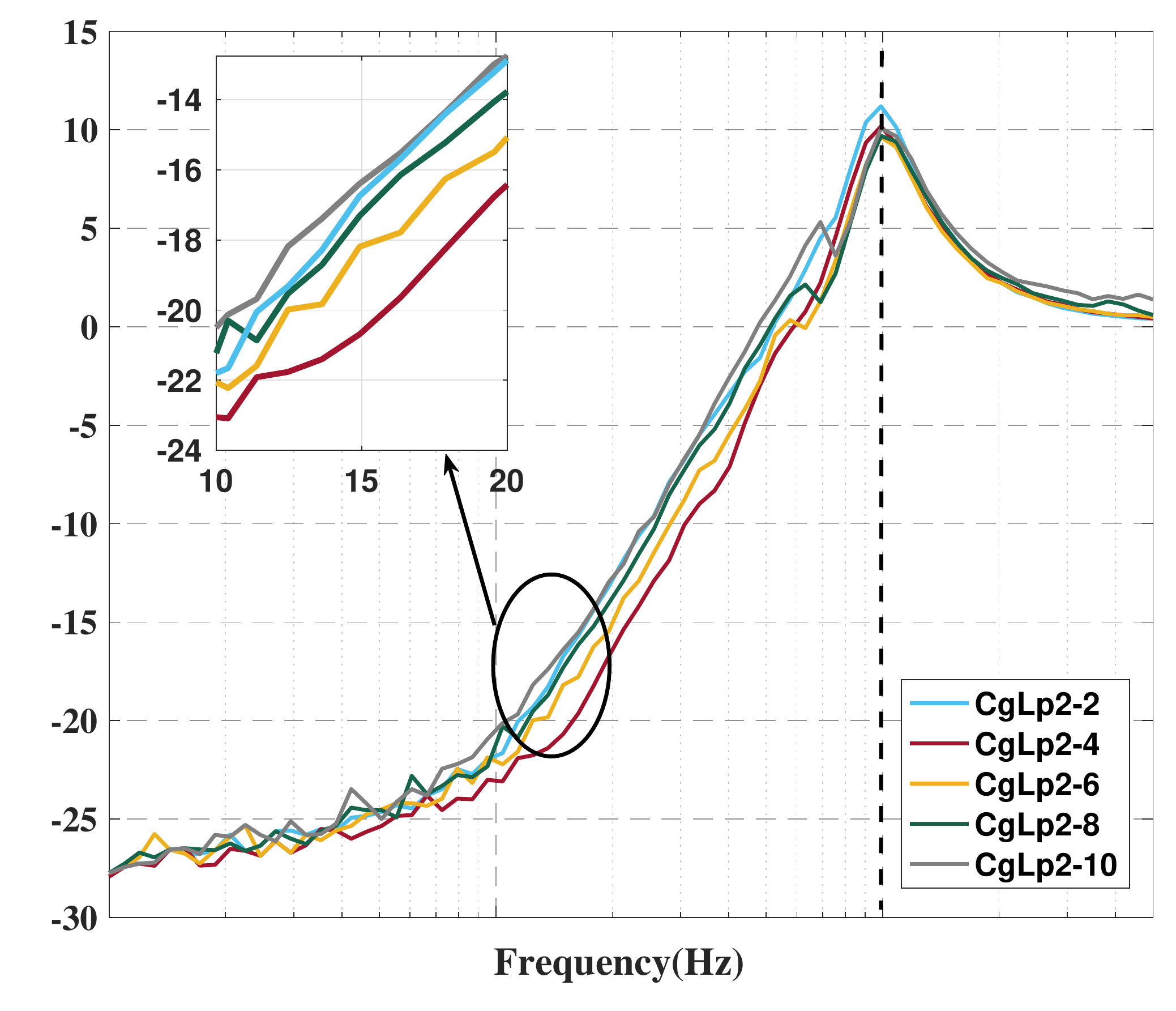}};
\draw (3.2,0.2) node [scale=0.6]  {$\omega_c$};
\draw (0.1,2) node [scale=0.6, rotate=90]  {$S_\infty$ (dB)};
\end{tikzpicture}
    \caption{The second order CgLp}
    \label{subfigb}
  \end{subfigure}  
  \caption{The defined pseudo-sensitivity $S_\infty$ of the mass system controlled by CgLp elements to achieve $30^\circ$ phase}  
  \label{Sensitivity}
\end{figure}

In this part, the closed-loop performance of a system controlled by CgLp configurations are analyzed through simulink (Fig. \ref{c1}). For simplicity, a mass system controlled by first and second order CgLp is considered (most of precision motion systems can be considered as a mass system). Additionally, effects of resonance are removed leading to easier analysis. The parameter $k_p$ is tuned so that all configurations have the cross-over frequency at 100 Hz.
The performance of the system is evaluated based on tracking precision using the defined pseudo-sensitivity $S_\infty$ (\ref{PS}). Since reference signals for tracking are composed of low frequency components in comparison to the cross-over frequency, $S_\infty$ behaviours are compared for frequencies smaller than $40Hz$ and the best configuration is selected. For this purpose, the designed controllers are digitalized with sampling time \SI{100}{\micro\second} using Tustin method \cite{1}. Then, the $r(t)=\sin(\omega t)$ is applied at each frequency while $n=0$ (Fig. \ref{c1}), and $S_\infty(\omega)$ is calculated through (\ref{PS}).
\newline For instance, $S_\infty(\omega)$ of configurations provide $30^\circ$ phase lead are shown in Fig. \ref{Sensitivity}. As was show, there are several CgLp compensators which provide $30^\circ$ phase. The configuration with lowest $S_\infty(\omega)$ is selected as the optimal configuration. Similarly, the optimal configurations for $20^\circ$, $40^\circ$ and $50^\circ$ compensation are obtained and all of the cases which provide optimal performance within their group are listed in Table \ref{foretuning}. Furthermore, to investigate the noise attenuation performance of these controllers, the reference signal is set as 0 and white noise (n) is added to the feedback branch of the system as shown in Fig. \ref{c1}. The configurations that have the best noise rejection are also shown in Table \ref{foretuning}. 

To understand how the open-loop higher order harmonics affect the closed-loop tracking performance, we characterize the $3^{rd}$ harmonic by $M_p$ and $\omega_p$. By comparison, it is found out that the configurations which has optimal tracking performance among the group always have the largest value of $\omega_p$ within the group. Also, these optimal configurations from a tracking perspective, have almost the lowest magnitude of third harmonics at low frequency among the group. In addition, the best configuration for noise attenuation in each group corresponds to the lowest $M_p$ in that group (the maximum value $\gamma$). These optimal cases have the lowest magnitude of harmonics at frequencies larger than $\omega_c$. These relations between open-loop $3^{rd}$ harmonic behaviour and closed-loop performance, although heuristic, provides us with insights for the tuning of CgLp elements. 
\newline Apart from relations in each group, for constant value of $\gamma$, since there is a direct relation between $\omega_{r\alpha}$ and $\omega_p$, reducing the $\omega_{r\alpha}$ will reduce the $\omega_p$ which causes low magnitude of higher order harmonics at high frequency. Thus, for a constant value of $\gamma$, reducing $\omega_{r\alpha}$ enhances noise rejection and increases phase margin at cross-over frequency; however, it sacrifices the tracking performance of the system.   
In conclusion, for a given phase compensation $\theta$, greater value of $\omega_p$ results in higher precision for tracking, and also, the lower value of $M_p$ results in the better noise rejection performance. Although the aforementioned results are obtained based on investigation for a mass system, results for mass spring damper systems with a low resonance frequency (as compared to control cross-over frequency) and a high damping coefficient are expected to show the same pattern. This is because CgLp controllers are usually used in combination with proportional integral (PI) controllers which shape the open-loop behaviour of considered mass spring damper systems into mass-like systems. Additionally, a high damping ratio ensures that the third harmonic is not higher than the first harmonic at any frequency. 
\begin{table*}[t!] 
\centering
\caption{RMS and maximum steady state error for noise attenuation and tracking performance (T and N indicate tracking and noise performance, respectively )}
\resizebox{\textwidth}{!}{
\begin{tabular}{|c|c|c|c|c|c|c|c|c|c|c|c|c|c|c|c|c|c|c|c|c|c|c|c|c|c|c|}
\hline
\multicolumn{3}{|c|}{\multirow{2}{*}{\textbf{$\gamma$}}}                   & \multicolumn{2}{c|}{-0.6} & \multicolumn{2}{c|}{-0.5} & \multicolumn{2}{c|}{-0.4} & \multicolumn{2}{c|}{-0.3} & \multicolumn{2}{c|}{-0.2} & \multicolumn{2}{c|}{-0.1} & \multicolumn{2}{c|}{0.0} & \multicolumn{2}{c|}{0.1} & \multicolumn{2}{c|}{0.2} & \multicolumn{2}{c|}{0.3} & \multicolumn{2}{c|}{0.4} & \multicolumn{2}{c|}{0.5} \\ \cline{4-27} 
\multicolumn{3}{|c|}{}                                                     & \textbf{N}  & \textbf{T}  & \textbf{N}  & \textbf{T}  & \textbf{N}  & \textbf{T}  & \textbf{N}  & \textbf{T}  & \textbf{N}  & \textbf{T}  & \textbf{N}  & \textbf{T}  & \textbf{N}  & \textbf{T} & \textbf{N}  & \textbf{T} & \textbf{N}  & \textbf{T} & \textbf{N}  & \textbf{T} & \textbf{N}  & \textbf{T} & \textbf{N}  & \textbf{T} \\ \hline
\multirow{6}{*}{$C_1$} & \multirow{2}{*}{$20^\circ$} & \textbf{Max $e(t)$} & 338         & 39          & 150         & 30          & 74          & 22          & 55          & 7           & 45          & 7           & 31          & 7           & 23          & 7          & 20          & 6          & 10          & 9          & 10          & 12         & -           & -          & -           & -          \\ \cline{3-27} 
                       &                             & \textbf{RMS}        & 143.6       & 15.5        & 43.9        & 13.1        & 32.8        & 9.9         & 29.3        & 1.9         & 14.5        & 2.1         & 19.9        & 1.7         & 7.2         & 2.1        & 7.8         & 1.7        & 3.6         & 2.1        & 3.3         & 4.63       & -           & -          & -           & -          \\ \cline{2-27} 
                       & \multirow{2}{*}{$30^\circ$} & \textbf{Max $e(t)$} & 65          & 27          & 63          & 10          & 53          & 7           & 35          & 7           & 29          & 7           & 26          & 7           & 13          & 8          & 8           & 13         & -           & -          & -           & -          & -           & -          & -           & -          \\ \cline{3-27} 
                       &                             & \textbf{RMS}        & 31.7        & 8.1         & 29.1        & 3.2         & 28.4        & 2.23        & 17.0        & 2.5         & 12.4        & 2.5         & 10.6        & 2.4         & 4.3         & 2.9        & 2.4         & 5.2        & -           & -          & -           & -          & -           & -          & -           & -          \\ \cline{2-27} 
                       & \multirow{2}{*}{$40^\circ$} & \textbf{Max $e(t)$} & 59          & 12          & 58          & 9           & 34          & 10          & 29          & 9           & 12          & 12          & 8           & 15          & 6           & 31         & -           & -          & -           & -          & -           & -          & -           & -          & -           & -          \\ \cline{3-27} 
                       &                             & \textbf{RMS}        & 30.3        & 4.2         & 30.1        & 2.8         & 14.1        & 3.2         & 13.4        & 3.0         & 3.3         & 4.3         & 2.6         & 5.7         & 2.0         & 12.5       & -           & -          & -           & -          & -           & -          & -           & -          & -           & -          \\ \hline
\multirow{6}{*}{$C_2$} & \multirow{2}{*}{$20^\circ$} & \textbf{Max $e(t)$} & -           & -           & -           & -           & 29          & 119         & 26          & 40          & 19          & 27          & 26          & 7           & 20          & 5          & 19          & 4          & 19          & 3          & 15          & 3          & 14          & 4          & 11          & 8          \\ \cline{3-27} 
                       &                             & \textbf{RMS}        & -           & -           & -           & -           & 9.8         & 69.2        & 6.7         & 18.0        & 5.4         & 13.45       & 4.9         & 1.9         & 10.0        & 1.4        & 7.5         & 1.2        & 5.4         & 1.2        & 4.4         & 1.2        & 4.0         & 13         & 3.3         & 2.9        \\ \cline{2-27} 
                       & \multirow{2}{*}{$30^\circ$} & \textbf{Max $e(t)$} & -           & -           & -           & -           & 20          & 38          & 20          & 17          & 19          & 8           & 19          & 6           & 16          & 5          & 16          & 5          & 14          & 4          & 11          & 4          & 10          & 5          & -           & -          \\ \cline{3-27} 
                       &                             & \textbf{RMS}        & -           & -           & -           & -           & 7.7         & 13.1        & 4.5         & 5.8         & 6.2         & 2.1         & 6.2         & 1.7         & 4.3         & 1.5        & 3.8         & 1.6        & 3.8         & 1.2        & 3.9         & 1.4        & 2.8         & 1.7        & -           & -          \\ \cline{2-27} 
                       & \multirow{2}{*}{$40^\circ$} & \textbf{Max $e(t)$} & -           & -           & -           & -           & 24          & 15          & 17          & 8           & 15          & 7           & 15          & 11          & 15          & 10         & 14          & 5          & 13          & 6          & 7           & 6          & -           & -          & -           & -          \\ \cline{3-27} 
                       &                             & \textbf{RMS}        & -           & -           & -           & -           & 7.4         & 4.8         & 6.4         & 2.3         & 5.4         & 2.2         & 4.6         & 3.7         & 4.8         & 3.7        & 4.4         & 1.2        & 5.2         & 2.1        & 2.5         & 2.1        & -           & -          & -           & -          \\ \hline
\end{tabular}
}
\label{noiseattenuation}
\end{table*}
\section{Validation}\label{sec:4}
\begin{figure}
\centering
\includegraphics[width=0.8\columnwidth]{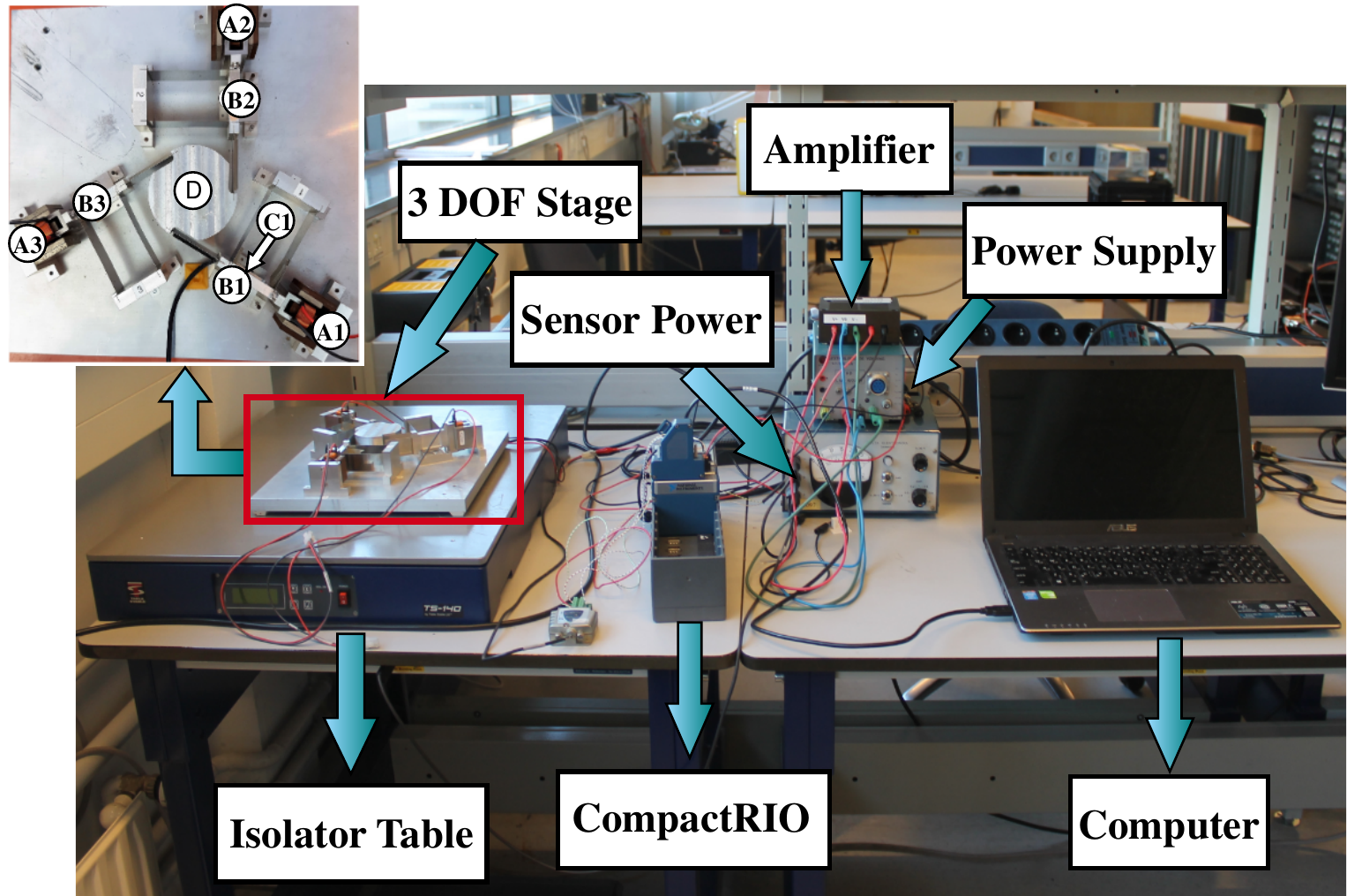}   
       \caption{The precision positioning stage}
    \label{setup}
\end{figure}

This section presents the experiments performed on a high precision positioning stage to validate the previously obtained tuning guidelines. This stage is a mass-spring-damper system which meets the condition of low resonance frequency compared to cross-over with a large damping factor. The precision positioning stage is shown in Fig. \ref{setup}. Three actuators are angularly spaced to actuate 3 masses (indicated by B1, B2 and B3) which are constrained by parallel flexures. These masses are connected to the central mass D through leaf flexures. For simplicity, only actuator A1 is utilized to control the position of B1, so we have a SISO system. Mercury M2000 encoder is used the measure the position of mass B1 with a resolution of \SI{100}{\nano\meter}. Figure \ref{identification} shows the frequency response of the stage. The system can be approximated as a single mass spring damper system with estimated transfer function as follows:
\begin{align}\label{transferfunction}
P _ { e s t } ( S ) = \frac {8695} {s ^ {2}+4.36s + 7627.3}
\end{align}
In order to validate the proposed relation, two control configurations  
$$
C _ {1}(s) = k _ { p } \underbrace { \left(1 + \frac { \omega _ { i } } { s } \right) } _ { \text {PI} }  CgLp_1\text{ and } C _ {2}(s) = k _ { p } \underbrace { \left(1 + \frac { \omega _ { i } }{s}\right)} _ { \text {PI} } CgLp_2
$$
are implemented on the precision positioning stage. The CgLp configurations of controllers are tuned to produce $20^\circ$, $30^\circ$, and $40^\circ$ degree phase lead at the crossover frequency. The cross-over frequency of all designed controllers is set to $100 Hz$. Also, $\omega_i$ is tuned as $\frac{\omega_c}{10}$ as per guideline provided in \cite{schmidt2014design}. All in all, we have six groups of control configurations and within each group the phase margin, the cross-over frequency and the type of CgLp element are the same.

The sinusoidal tracking experiments are carried out to validate the tuning guidelines based on the defined pseudo-sensitivity $S_\infty$ analysis. Since it is time-consuming to obtain the behaviour of $S_\infty$ over the entire frequency range, the maximum steady state errors of sinusoidal inputs with frequencies of $5Hz$, $10Hz$, and $20Hz$ were used for performance analysis. Figure \ref{FORECGLP} shows the maximum steady state error of reference tracking and corresponding $\omega_p$ for each control configuration of the first and second order CgLp. It is seen from the figures that the lowest value of maximum steady state error is obtained when the  $\omega_p$ is the maximum which is consistent with the analysis in the previous section.

In the experiments of noise rejection, zero reference is used and additional white noise with a maximum magnitude of $5000$ nm (50 times the resolution of the sensor) is applied to the feedback branch of the system as shown in Fig. \ref{c1}, and the results are shown in Table \ref{noiseattenuation}. It can be seen that the configurations with optimal noise attenuation performance within the group are consistent with cases indicated in Table \ref{foretuning}. Moreover, it is noteworthy that the optimal cases of first order CgLp outperform that of second order CgLp for noise attenuation. This can be explained by the fact that the optimal cases of the first order CgLp have the lowest magnitudes of $M_p$ than the ones of the second order as seen in Table \ref{foretuning}.  
\begin{figure}
\centering
     \includegraphics[width=0.8\linewidth]{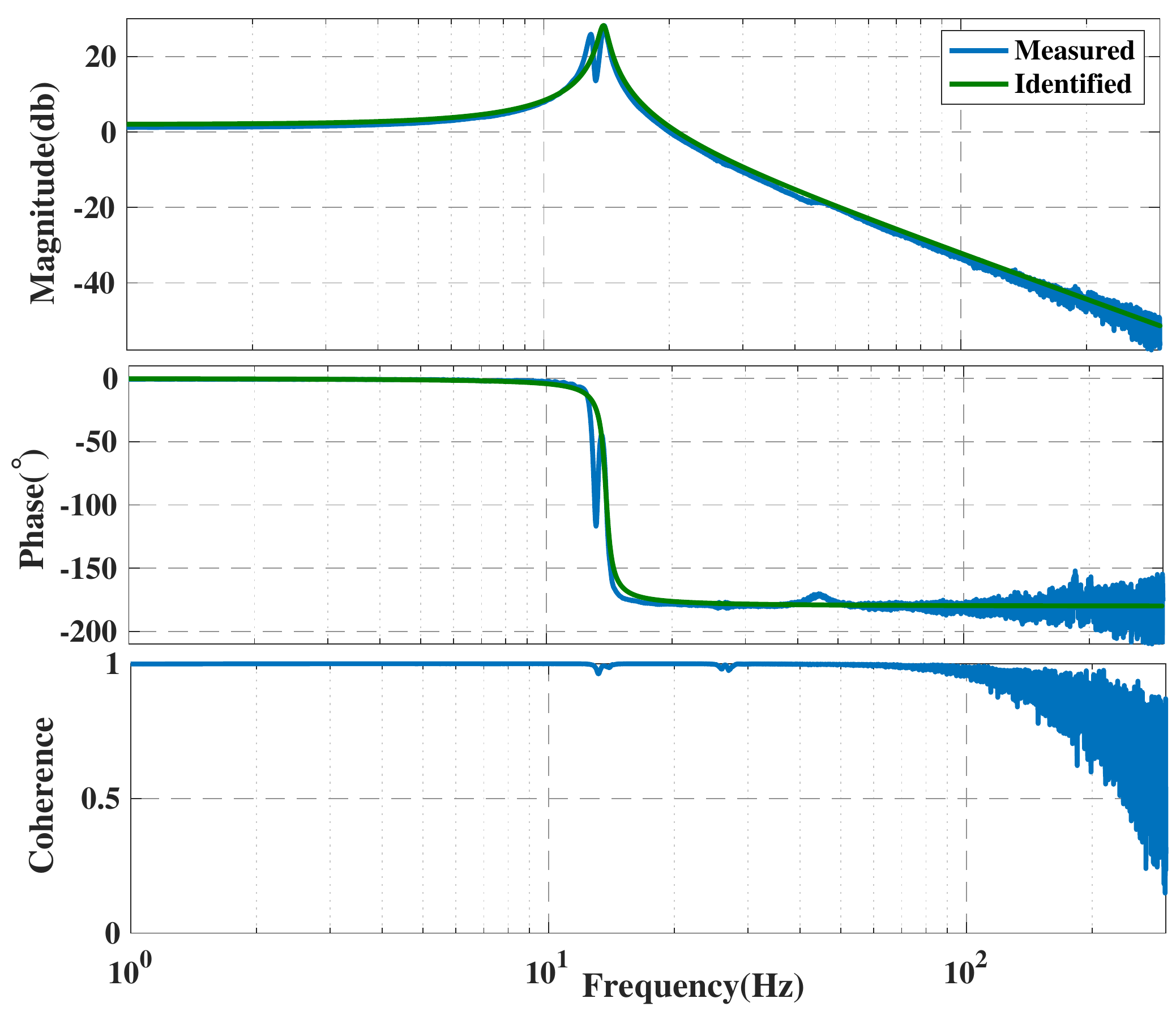}
    \caption{Identification of the precision positioning stage}
    \label{identification}
\end{figure}
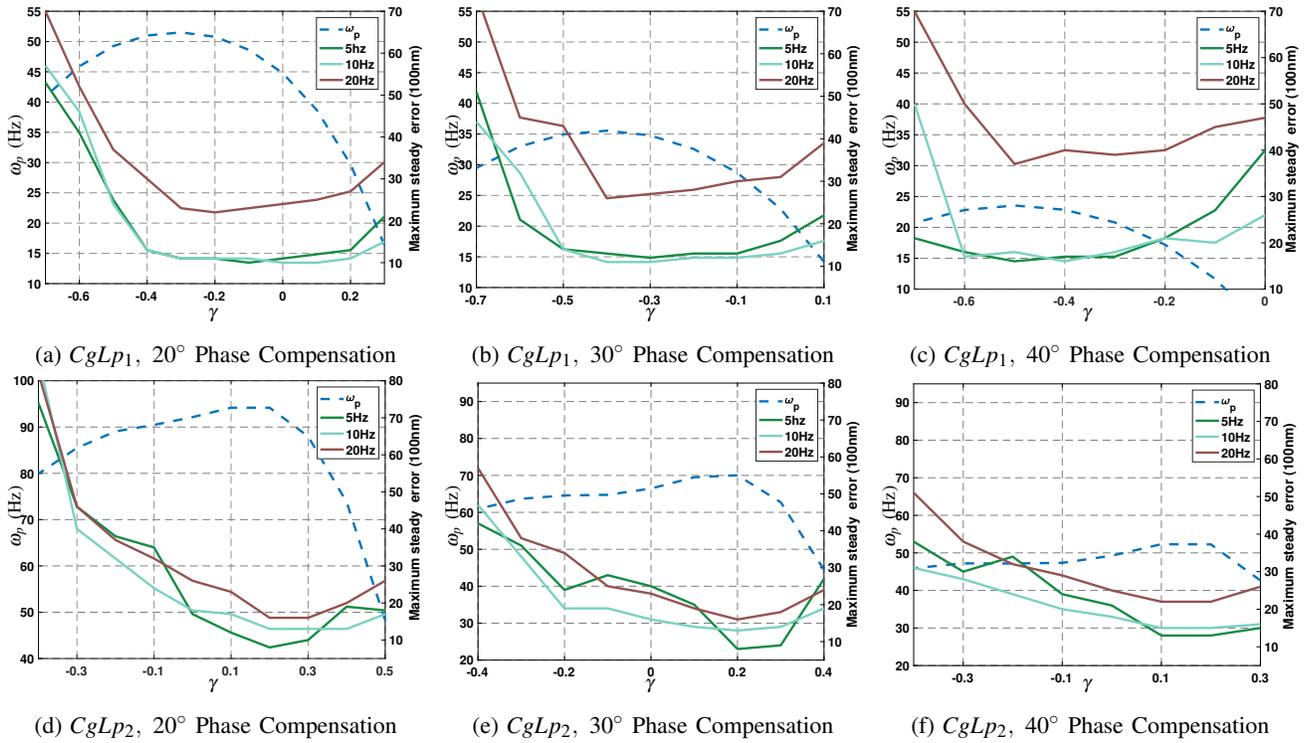
\begin{figure*}
  \centering
  \begin{subfigure}{0.32\linewidth}
    \centering
    \begin{tikzpicture}
    \node[anchor=south west,inner sep=0] at (0,0) {\includegraphics[width=\linewidth]{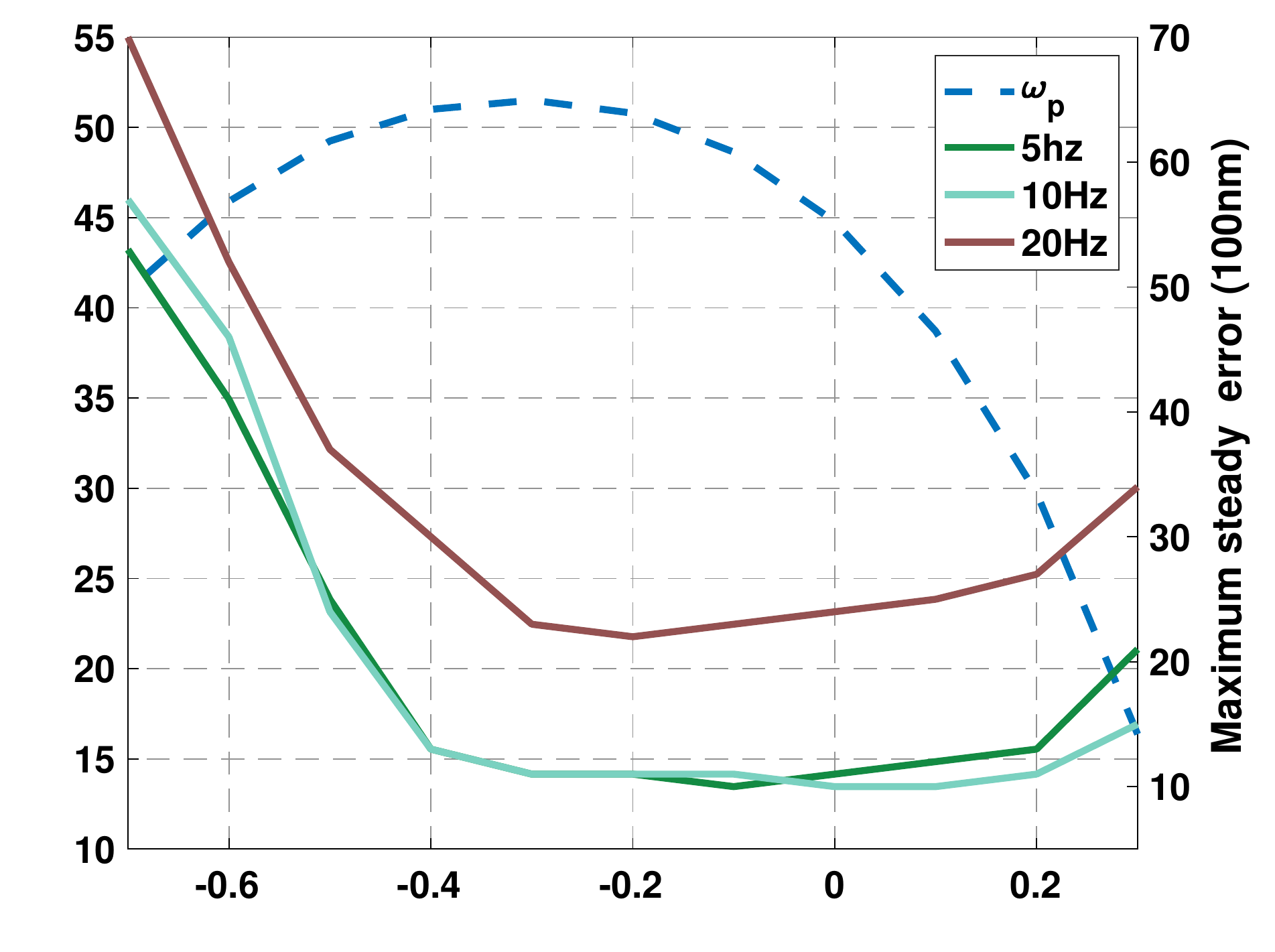}};
\draw (2.8,0.05) node [scale=0.7]  {$\gamma$};
\draw (0.2,2.3) node [scale=0.7, rotate=90]  {$\omega_p$ (Hz)};
\end{tikzpicture}
    \caption{$CgLp_1,\ 20^\circ$ Phase Compensation}
  \end{subfigure}
\begin{subfigure}{0.32\linewidth}
    \centering
    \begin{tikzpicture}
    \node[anchor=south west,inner sep=0] at (0,0) {\includegraphics[width=\linewidth]{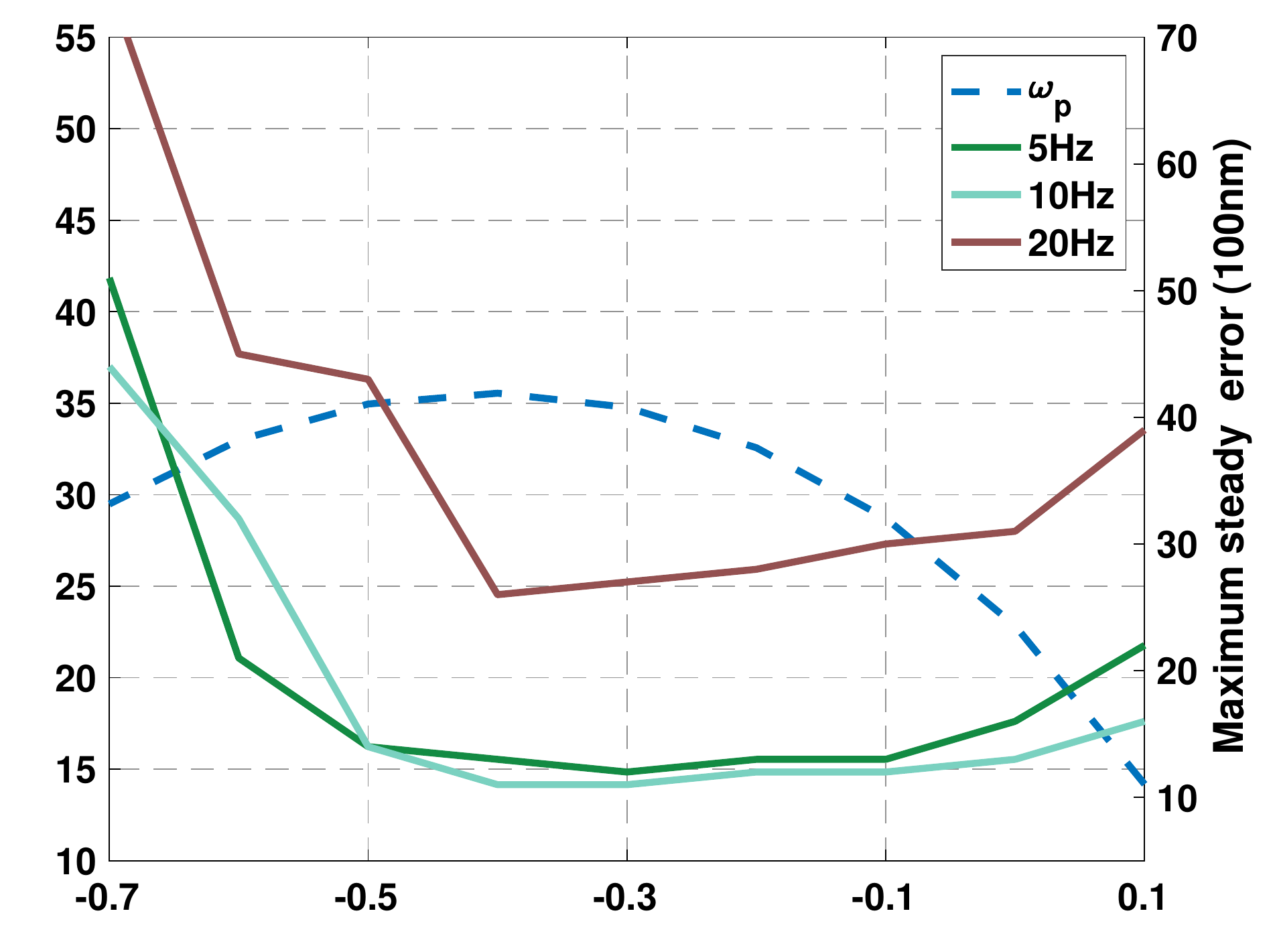}};
\draw (2.8,0.05) node [scale=0.7]  {$\gamma$};
\draw (0.2,2.3) node [scale=0.7, rotate=90]  {$\omega_p$ (Hz)};
\end{tikzpicture}
    \caption{$CgLp_1,\ 30^\circ$ Phase Compensation}
  \end{subfigure}  
  \begin{subfigure}{0.32\linewidth}
    \centering
    \begin{tikzpicture}
    \node[anchor=south west,inner sep=0] at (0,0) {\includegraphics[width=\linewidth]{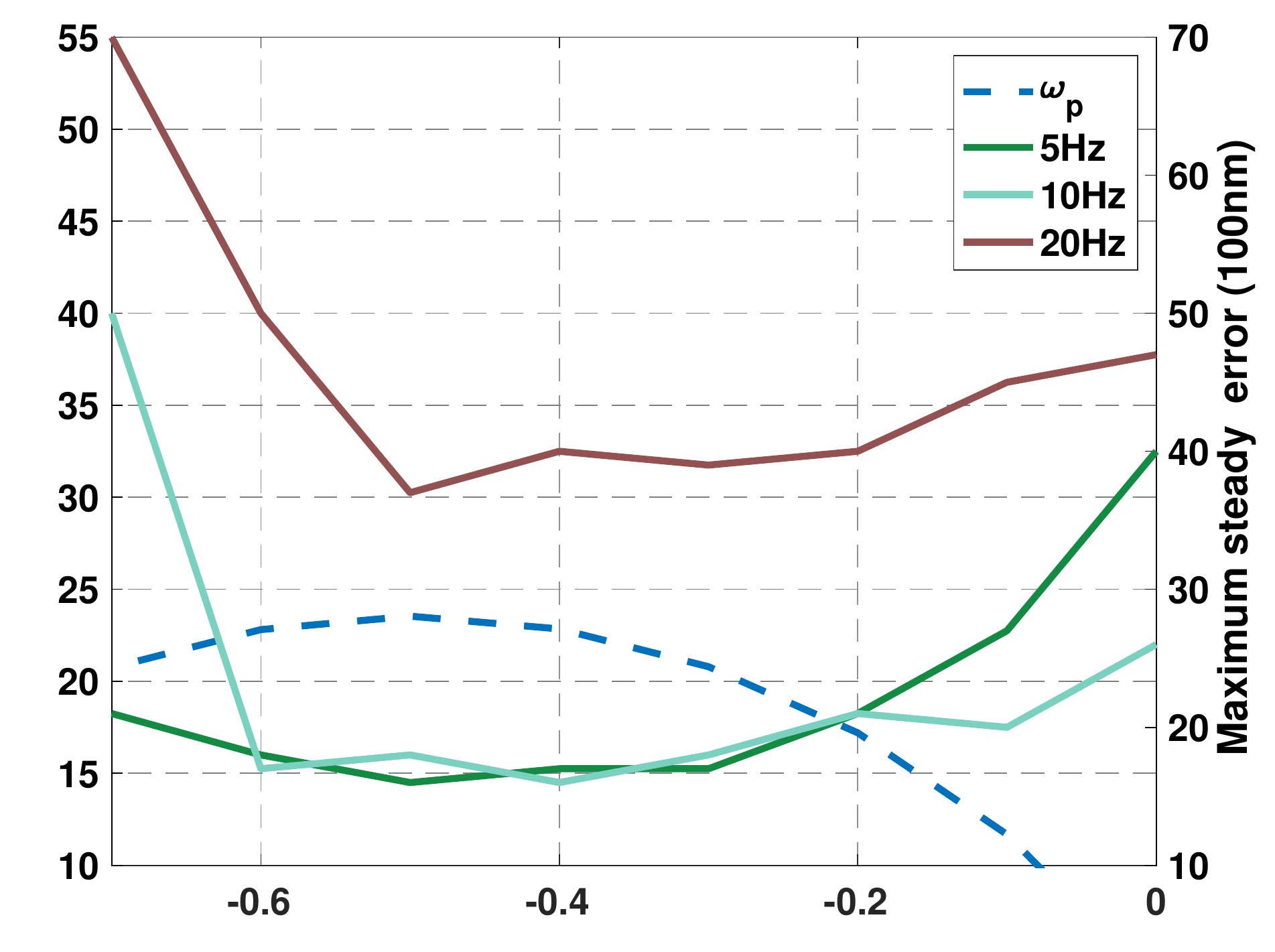}};
\draw (2.8,0.05) node [scale=0.7]  {$\gamma$};
\draw (0.2,2.3) node [scale=0.7, rotate=90]  {$\omega_p$ (Hz)};
\end{tikzpicture}
    \caption{$CgLp_1,\ 40^\circ$ Phase Compensation}
  \end{subfigure}  
\begin{subfigure}{0.32\linewidth}
    \centering
    \begin{tikzpicture}
    \node[anchor=south west,inner sep=0] at (0,0) {\includegraphics[width=\linewidth]{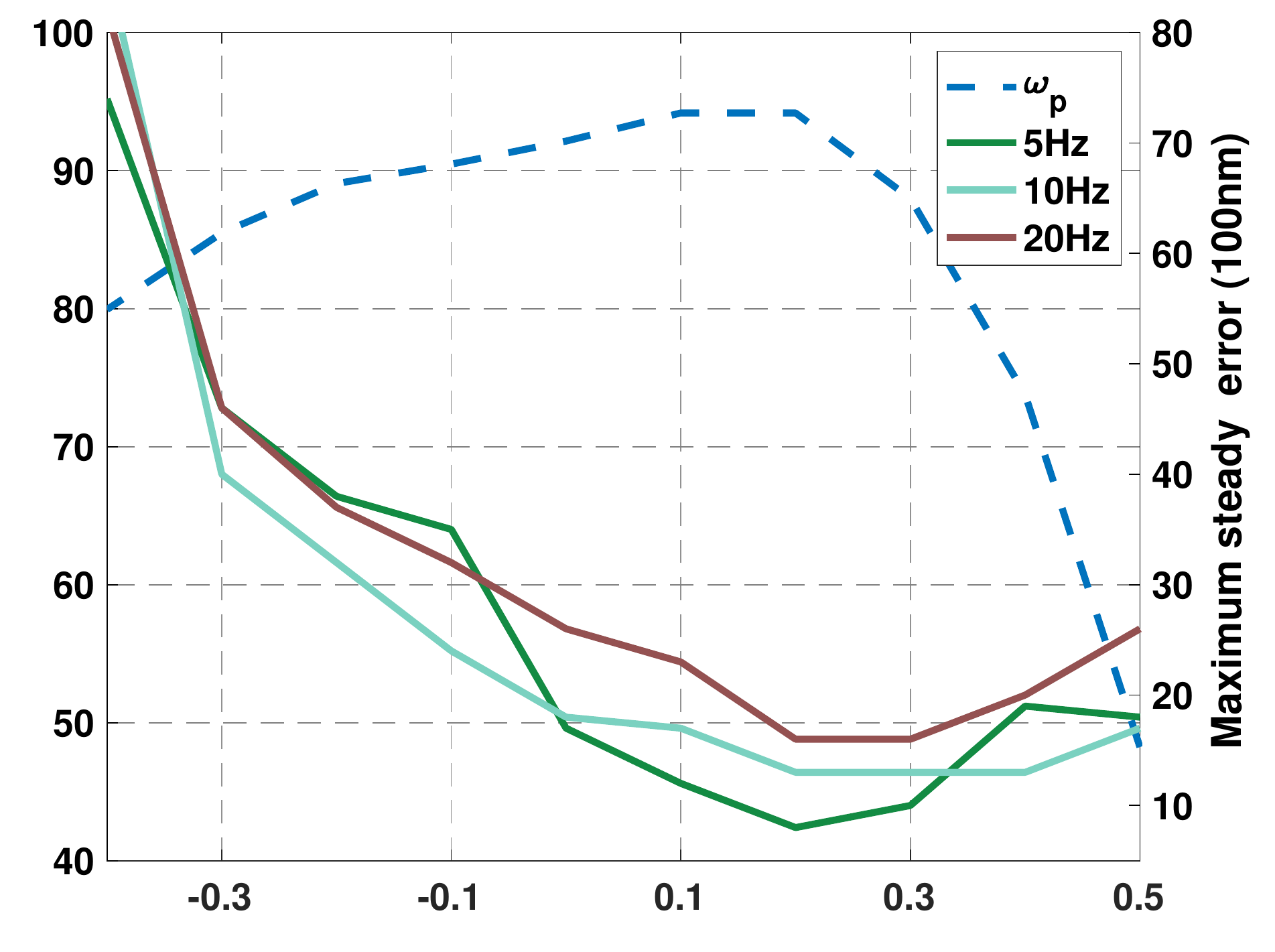}};
\draw (2.8,0.05) node [scale=0.7]  {$\gamma$};
\draw (0.2,2.3) node [scale=0.7, rotate=90]  {$\omega_p$ (Hz)};
\end{tikzpicture}
    \caption{$CgLp_2,\ 20^\circ$ Phase Compensation}
  \end{subfigure}  
\begin{subfigure}{0.32\linewidth}
    \centering
    \begin{tikzpicture}
    \node[anchor=south west,inner sep=0] at (0,0) {\includegraphics[width=\linewidth]{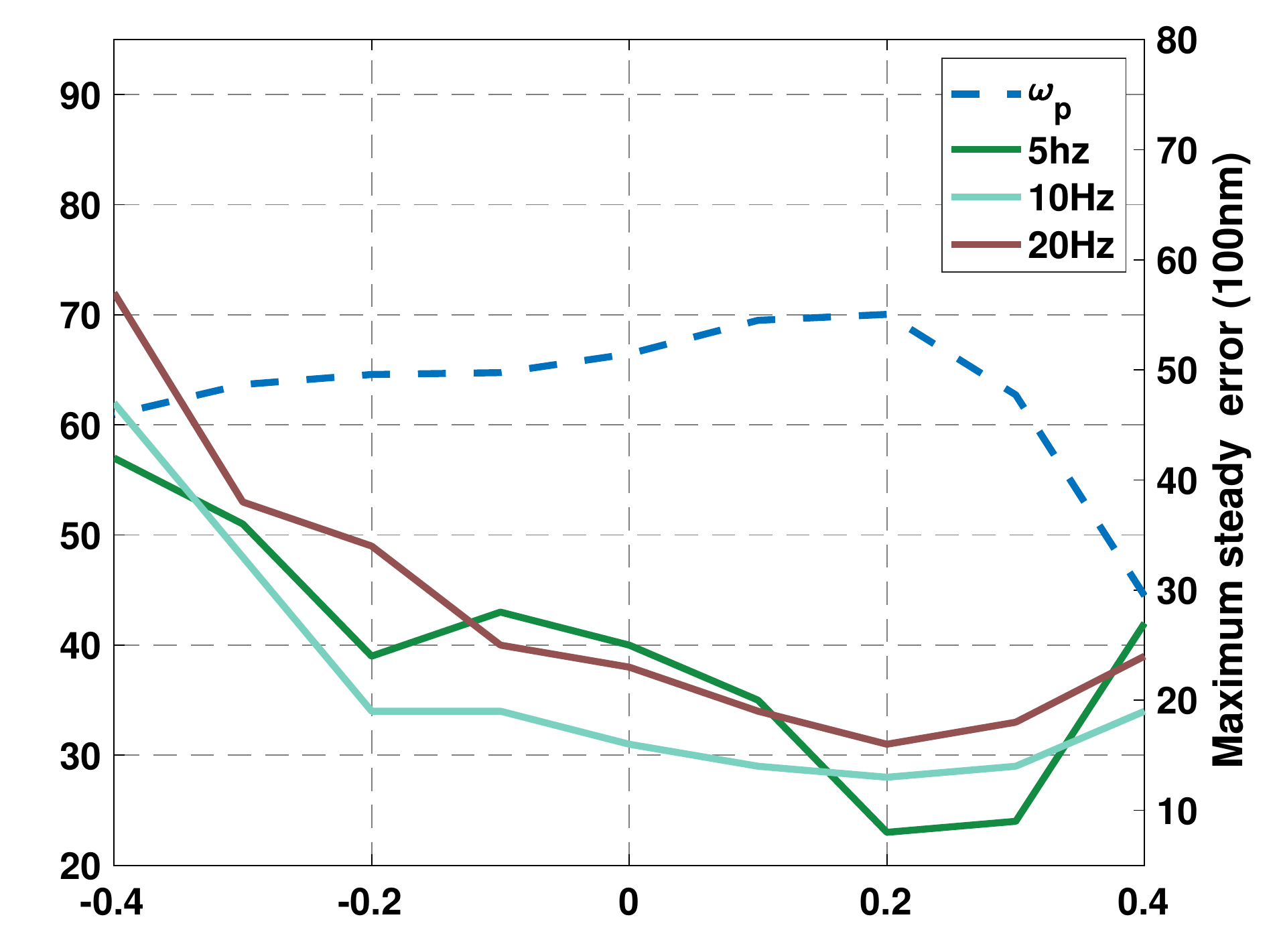}};
\draw (2.8,0.05) node [scale=0.7]  {$\gamma$};
\draw (0.2,2.3) node [scale=0.7, rotate=90]  {$\omega_p$ (Hz)};
\end{tikzpicture}
    \caption{$CgLp_2,\ 30^\circ$ Phase Compensation}
  \end{subfigure}  
\begin{subfigure}{0.32\linewidth}
    \centering
    \begin{tikzpicture}
    \node[anchor=south west,inner sep=0] at (0,0) {\includegraphics[width=\linewidth]{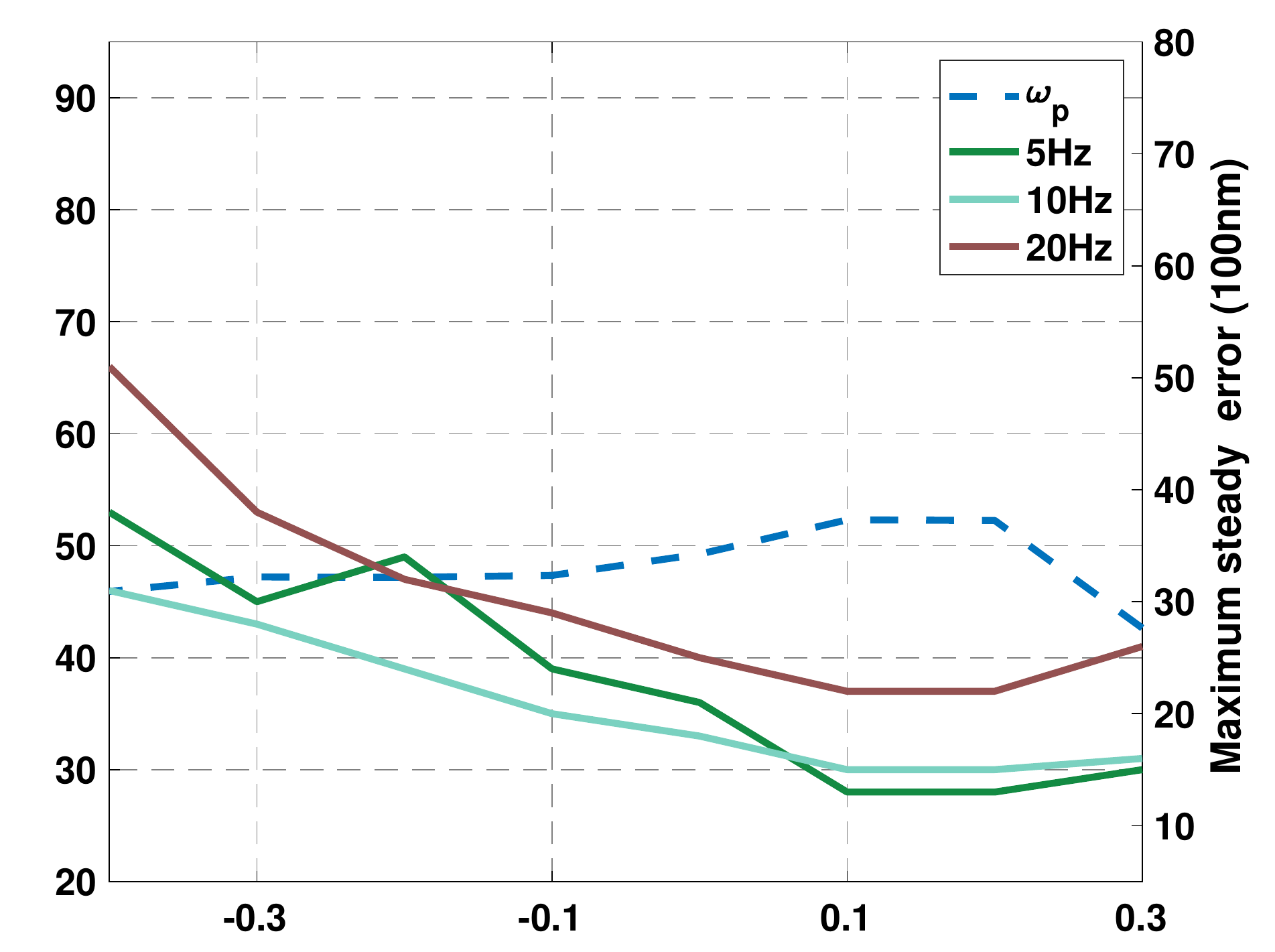}};
\draw (2.8,0.05) node [scale=0.7]  {$\gamma$};
\draw (0.2,2.3) node [scale=0.7, rotate=90]  {$\omega_p$ (Hz)};
\end{tikzpicture}
    \caption{$CgLp_2,\ 40^\circ$ Phase Compensation}
  \end{subfigure}  
  \caption{Maximum steady state error of the system with various sinusoidal reference inputs}  
  \label{FORECGLP}
\end{figure*} 
\begin{figure}
    \centering
    \includegraphics[width=0.65\columnwidth]{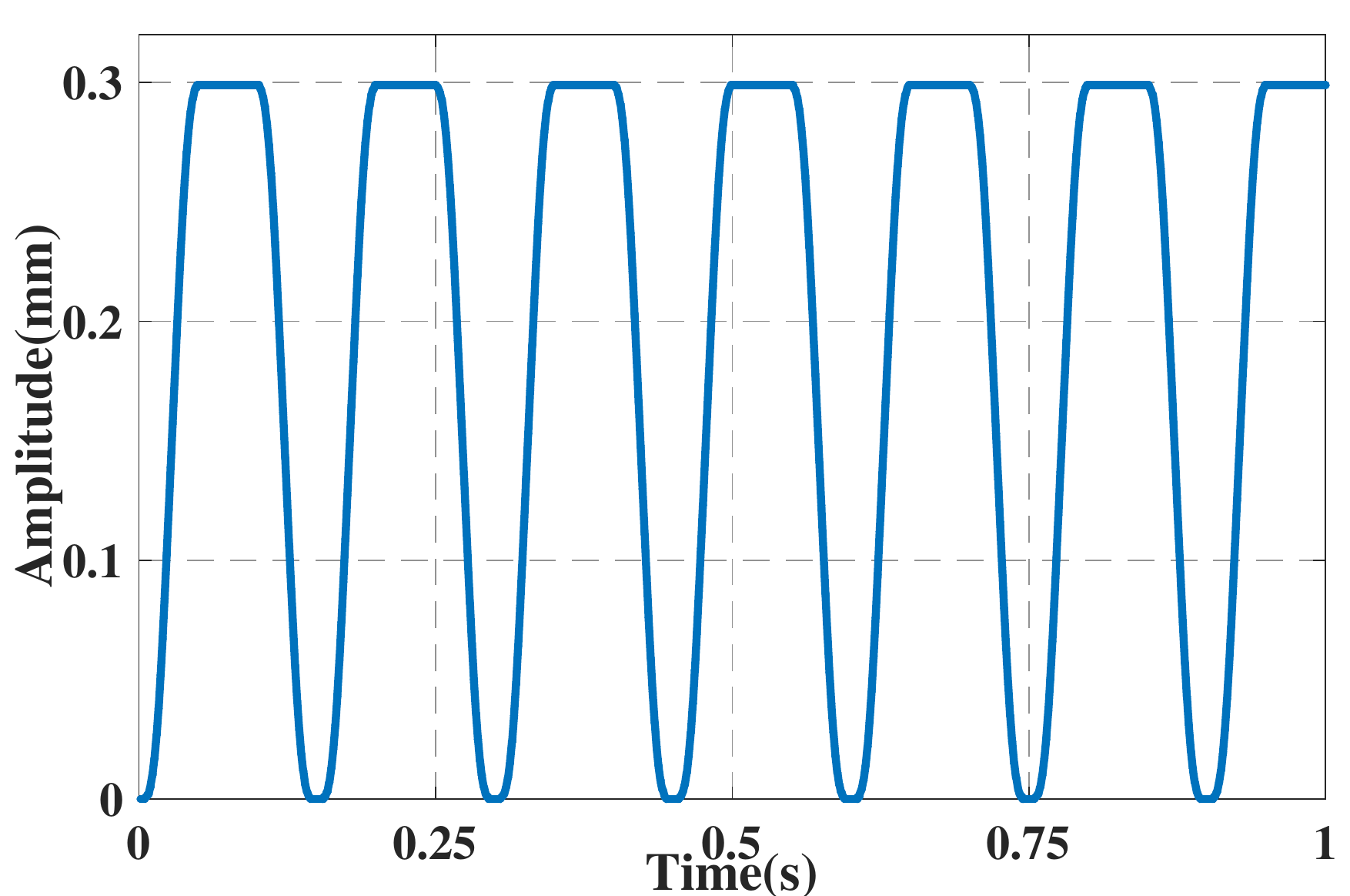}
    \caption{Trajectory}
    \label{trajectory}
\end{figure}

Furthermore, since industrial motion control is achieved with a predefined trajectory, the tracking performance for a smooth trajectory (Fig. \ref{trajectory}) which is the combination of sinusoidal waves with different frequencies is investigated. Table \ref{noiseattenuation} shows the Root Mean Square (RMS) and maximum steady state error for each scenario. It is observed from the tables that the optimal performance regarding RMS and maximum steady state error are still obtained with reset values that produce a maximum $\omega_p$. Additionally, from the results, we see that the optimal case in the sense of tracking performance has the moderate performance in the sense of noise rejection. Indeed while CgLp has allowed us to reduce waterbed effects of linear controllers, a new design trade off between noise rejection and tracking performance is created by higher order harmonics for this kind of controllers which has been found in this paper. Based on these analyses, the tuning method can be summarized as: 
\begin{enumerate}
  \item Use describing function to design a group of CgLp elements that provide the required phase lead as compensation at crossover frequency.
  \item  Check the frequency $\omega_p$ at which the $3^{rd}$ harmonic peak is placed for each configuration using HOSIDF. 
  \item  If tracking performance is important, choose the CgLp configuration that has the largest value of $\omega_p$.
  \item If noise rejection is important, choose the configuration with lowest $M_p$ (highest value of $\gamma$).
  \end{enumerate}
\section{Conclusion}\label{sec:5}
This paper has proposed a tuning guideline for Constant in gain Lead in phase (CgLp) configurations. Several groups of CgLp controllers are designed to achieve the same amount of phase compensation, and their open-loop higher order harmonics behaviour are investigated through higher order sinusoidal describing function analysis (HOSIDF). Then, the closed-loop tracking precision performances are evaluated by a defined pseudo-sensitivity function which considers all harmonics. It is found that the optimal tracking precision performances are obtained with cases that have the largest frequency of $3^{rd}$ harmonic peak ($\omega_p$) which have almost the smallest magnitude of higher order harmonics at low frequencies. On the other hand, configurations with maximum value of $\gamma$ (the lowest value of $M_p$) which result in the lowest magnitude of $3^{rd}$ harmonic at higher frequencies have the best noise attenuation performance. Results are also validated by the experiments performed on a precision positioning stage. Although the exact relation between open-loop higher order harmonics and closed-loop performance has not completely been theoretically established, this paper provides guidelines to minimize the negative effect of higher order harmonics on the performance of the system. For future study, it is interesting to establish a $H_\infty$ by combining it with the proposed tuning guideline.    
 \bibliographystyle{IEEEtran}
 \bibliography{reference}
\end{document}